\documentclass[aps,prl,showpacs,twocolumn, superscriptaddress,groupedaddress,reprint]{revtex4}

\usepackage{subfigure}
\usepackage{wrapfig}
\usepackage{textcomp}
\usepackage{graphicx}
\usepackage{amssymb}
\usepackage{amsmath}
\usepackage{bm}
\usepackage{amsmath, amsthm, amssymb}
\usepackage{hyperref}
\usepackage{color}
\usepackage{rotating}

\usepackage{amsfonts}
\usepackage{dcolumn}
\usepackage{float}
\usepackage{bm}

\begin{document}

\title{Mixed Stochastic-Deterministic Time-Dependent Density Functional Theory: Application to Stopping Power of Warm Dense Carbon}

\author{Alexander J. White$^{1}$, Lee A. Collins$^{1}$, Katarina Nichols$^{1,2}$, and S. X. Hu$^{2}$}
\address{alwhite@lanl.gov, 1Theoretical Division, Los Alamos National Laboratory, Los Alamos, 87545 NM,  USA, 2 University of Rochester, Laboratory of Laser Energetics, Rochester 14623 NY, USA}

\date{\today}

\begin{abstract}
Warm dense matter (WMD) describes an intermediate phase, between condensed matter and classical plasmas, found in natural and man-made systems. In a laboratory setting, WDM needs to be created dynamically. It is typically laser or pulse-power generated and can be difficult to characterize experimentally. Measuring the energy loss of high energy ions, caused by a WDM target, is both a promising diagnostic and of fundamental importance to inertial confinement fusion research. However, electron coupling, degeneracy, and quantum effects limit the accuracy of easily calculable kinetic models for stopping power, while high temperatures make the traditional tools of condensed matter, \emph{e.g.} Time-Dependent Density Functional Theory (TD-DFT), often intractable. We have developed a mixed stochastic-deterministic approach to TD-DFT which provides more efficient computation while maintaining the required precision for model discrimination. Recently, this approach showed significant improvement compared to models when compared to experimental energy loss measurements in WDM carbon. Here, we describe this approach and demonstrate its application to warm dense carbon stopping across a range of projectile velocities. We compare direct stopping-power calculation to approaches based on combining homogeneous electron gas response with bound electrons, with parameters extracted from our TD-DFT calculations. \end{abstract}

\maketitle

\section{Introduction}
	Warm Dense matter is both found in astrophysical systems \cite{Ehrenreich_2020,REDMER2011798,Nettelmann_2008}, ranging from planetary interiors to solar cores \cite{Benuzzi_Mounaix_2014,Guillot:1999aa,Nguyen_2004,Stevenson_2009,Chabrier_2000,10.2307/2892796}, and generated in dynamic high energy density experiments \cite{Falk_2018,doi:10.1063/5.0007476,PhysRevLett.90.175002}, \emph{e.g.} using Xray free-electron lasers (XFELs) \cite{doi:10.1063/5.0048150} or in inertial confinement fusion (ICF) shots \cite{Hinkel_2013}. One method of both generation \cite{doi:10.1063/5.0026595, PhysRevE.100.043204} and  characterization\cite{Hayes:2020un, doi:10.1063/1.4928104} of WDM is with high energy ion beams. High energy ($k$eV-$M$eV) particles (projectiles), often protons or $\alpha$-particles, are shot into a target, heating electrons and potentially generating WDM.\cite{PhysRevE.92.063101} For ICF, ion stopping is critical to topics such as $\alpha$-particle self heating \cite{doi:10.1063/1.5030337}, ion-driven fast ignition \cite{Fern_ndez_2009}, or heavy-ion fusion \cite{RevModPhys.86.317,HOFMANN20181,doi:10.1063/1.874031}. Measuring the loss of energy of the ion beam as it passes through a finite thickness target allows for the calculation of the target's electronic stopping power (ESP). Combining this ion-beam measurement with laser generation of WDM allows one to measure the ESP of WDM directly. \cite{PhysRevLett.114.215002, PhysRevLett.115.205001, PPR:PPR393148}  For velocities near the electron velocity, $v_{e}$ the ESP becomes sensitive to the density and temperature ($\rho$ \& T) of the system, allowing for a powerful diagnostic \cite{PhysRevLett.115.205001}. However, inference of $ \rho$ and/or T directly from energy loss measurements requires a predictive ESP model \cite{Cayzac:2017un}.  Models are often in agreement for classical plasmas or  high-velocity projectiles $v_{p}>>v_{e}$, where the electron-projectile interaction is weak. However, there is significant disagreement for velocities near the so-called `Bragg Peak', \emph{ i.e.} $v_{p}\approx v_{e}$ where the interaction is strongest. \cite{PhysRevE.92.053109,Cayzac:2017un}
	
\begin{figure}[t]
	\begin{center}
	\includegraphics[width=1.0\columnwidth] {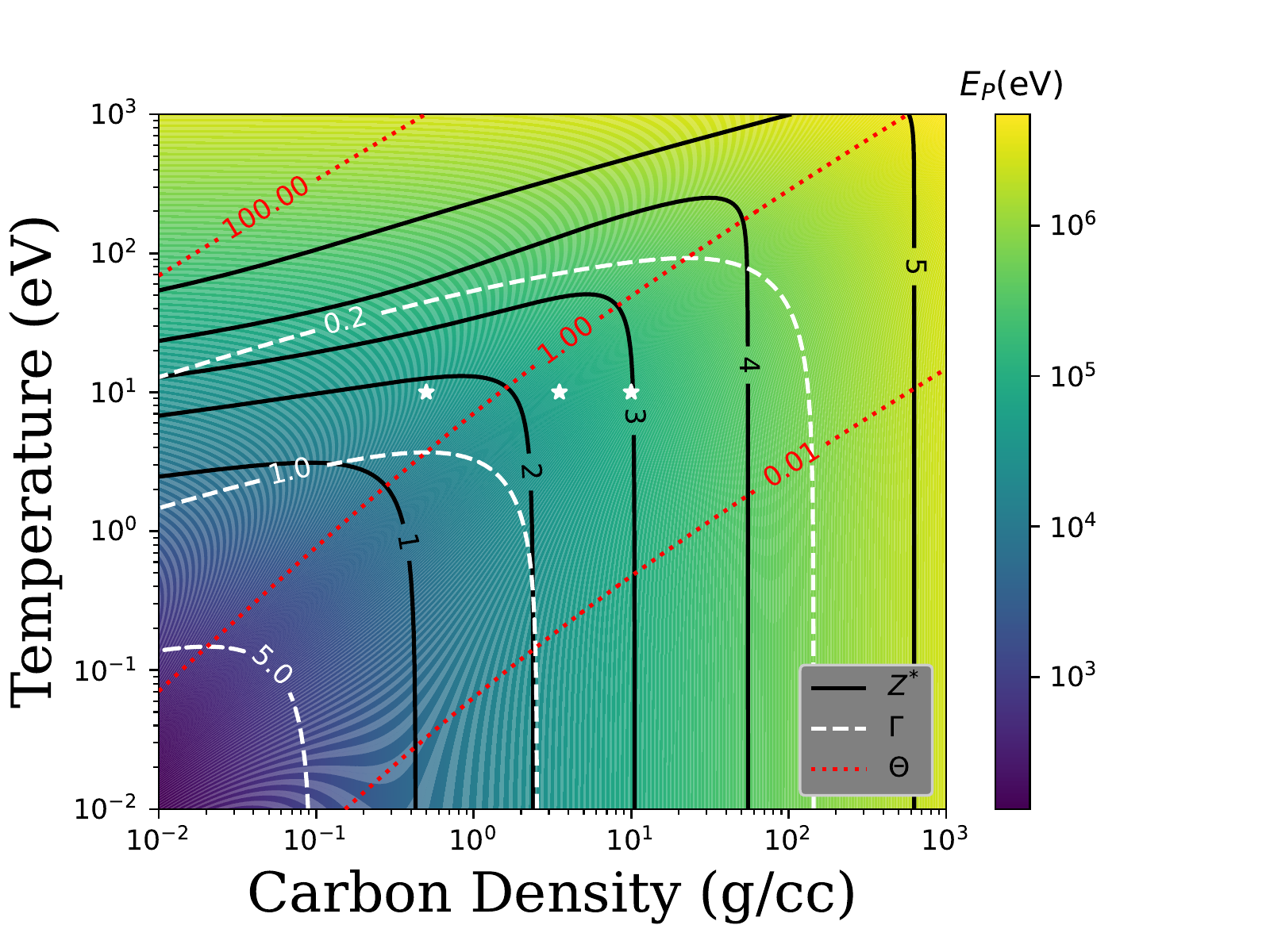}
	\caption{(Color online) Electron kinetic energy, $K_e$, for Carbon based on Thomas Fermi approximation. Colorbar shown on right. Solid contours show the integer ionization states,  white-dashed and red-dotted contours show the electron degeneracy $\Theta$ and electron coupling $\Gamma$ across 4 orders of magnitude. White stars indicate the 3  density \& temperatures explicitly considered in this article. }
	\label{plasma_param}
	\end{center}
\end{figure}

	Modeling of WDM, and ESP of WDM,  is complicated by the combination of non-negligible electronic coulomb interaction, degeneracy and thermal excitation. That is, the (partially degenerate) plasma coupling parameter, $\Gamma={E_{C}\over{T + E_F}}$, the electron degeneracy, $\Theta={T\over{E_{F}}}$, and their inverses are non-negligible.  $E_C \sim r_{WS}$ is the Coulomb energy, $r_{WS}$ is the Wigner Seitz radius, $E_F=1/2(3\pi^2\rho_e)^{2/3}$ is the Fermi energy, and $\rho_e$ is the electron number density.   With the exception of identifying the $\rho$ and $T$ of our carbon systems, we utilize Hartree atomic units throughout the article, \emph{i.e.} $\hbar$, $m_e$, $e$, $k_B$, and $4\pi\epsilon_0 = 1$, including all figures, parameters, and equations. Figure \ref{plasma_param} shows these parameters, $\theta$ red dotted contour, $\Gamma$ white dashed contour, for the carbon ``free" electron contribution, based on the Thomas Fermi model for the partial ionization \cite{doi:10.1063/1.871806}, $Z^*$ black solid contour. 

For warm dense carbon, and other mid-to-high atomic number ($Z$) materials, the changing $Z^*$ with $\rho$ \& T adds additional complication. $Z^*$ is a poorly defined quantity for atoms in dense phases. Though they can be useful in roughly determining the state of a system, plasma parameters such as $v_e$, $\Gamma$, and $\Theta$ are only strictly valid in nearly-homogeneous electron gas, where a single $E_F$ is a meaningful quantity. The nearly-homogeneous 'free' electron density, depends on the poorly defined $Z^*$.  
	
	The pure plasma picture is overly simplistic for a material such as carbon where inner shell (K-shell), outer shell (L-shell) and different angular momenta (s, p) electrons can contribute to material response to varying degrees. While energetic separation between K and L shell may be large enough to allow for clear distinction, in dense systems the intra-shell ionization of the L-shell is model and definition dependent. \emph{Ab initio} methods, such as Density Functional Theory (DFT), do not depend on a model for defining a sharp separation between ``core" and ``free"  electrons, at least for the outer shells.
	 	
	Due to their similar velocities, high energy ($k$eV-$M$eV) projectile ions interact primarily with the electrons, rather than bulk ions. The average force against the projectile motion is known as the stopping power of that bulk material for the type of projectile. The interaction is greatest when the electron and projectile velocities are close, \emph{i.e.} when $v_p \approx v_e$. Here we have defined 
\begin{eqnarray}
\label{ve}
	v_{e}\equiv\sqrt{3T+ v^2_{F}}\:, 
\end{eqnarray}	
	where $v_{F}$ is the Fermi velocity. This definition accounts for both thermal excitation and degeneracy. The colormap shown in Fig. \ref{plasma_param} shows the proton projectile energy, $E_p$ in eV, where $v_p \approx v_e$, \emph{ i.e.} the estimated `Bragg Peak' position for a proton projectile.

	 Analytical models typically coalesce in the  limit of high projectile velocities, assuming they use the same plasma conditions, \emph{i.e.} $Z^*$. However, for lower projectile velocities near the Bragg peak, the energy and wavevector dependence of the materials response functions becomes important and these models disagree. This is also the regime where the ESP becomes sensitive to the temperature of the WDM target and shows non-trivial density dependence \cite{White:2018aa}. Thus it is the ideal regime for plasma diagnostic measurements. Often, the plasma conditions of WDM target can only be estimated from hydrodynamic calculations.\cite{Duffy_2006,PPR:PPR393148} One needs an accurate atomistic and \emph{i.e. ab initio}, model such as time-dependent density functional theory (TD-DFT), in order to validate these more approximate methods, or to provide an alternative method for comparison to experimental measured stopping.\cite{PhysRevLett.121.145001,White:2018aa}

	For ambient conditions, direct simulation of ESP through TD-DFT, using periodic boundary conditions, has arisen as a high accuracy method for calculation of ESP \cite{PhysRevB.91.014306,Sand:2019wi}. However, for WDM its application is limited due to a combination of finite simulation-box effects and increased computational cost due to high temperatures. This lead us to the development of orbital-free TD-DFT methods\cite{PhysRevLett.121.145001,White:2018aa}, which proved accurate at high velocities but are insufficient for velocities near the Bragg Peak \cite{White:2018aa}. Fortunately, the development of stochastic DFT / TD-DFT (sDFT / TD-sDFT) provide a feasible alternative to orbital-free TD-DFT or analytical models, while retaining the full accuracy of Kohn-Sham (KS) DFT \cite{Cytter:2018aa,doi:10.1002/wcms.1412}. However this comes at the cost of finite precision due to stochastic error. The sDFT calculation of initial electron density scales as $V/T$ where $V$ is the system size and $T$ is the temperature, while the proceeding TD-sDFT calculation only scales as $V$\cite{Cytter:2018aa}.  The precision of TD-sDFT scales as $1/\sqrt{N_\chi}$ where $N_\chi$ is the number of `stochastic vectors', $\chi$. Achieving desirable stochastic error often requires large numbers of orbitals to converge \cite{Chen_2019,Chen_2019_2,Li_2019,Neuhauser_2014}. This is improved by generalizing our recently developed mixed stochastic-deterministic algorithm \cite{PhysRevLett.125.055002}, mDFT, to the novel TD-mDFT for time-dependent systems. We demonstrate this new method on simulations of warm dense carbon. A subset of these simulations were recently validated by experiment energy-loss measurements.\cite{PPR:PPR393148} 
	
	In this article we will 1) briefly discuss the traditional models for ESP, 2) discuss the use of TD-DFT to calculate ESP at an atomistic level, 3) briefly review stochastic TD-DFT, 4) describe our mixed stochastic-deterministic TD-DFT method, and 5) present results for the stopping power of carbon at 0.5, 3.5, and 10 g/cc and a temperature of 10 eV, with comparison to dielectric models with bound-electron corrections and orbital-free TD-OF-DFT. Finally, we will attempt to combine atomistic calculation of electron relaxation times and $Z^*$, from electrical conductivity calculations or electronic density of states to improve agreement between the model and atomistic TD-DFT. 

\section{Theory}
\subsection{Stopping Power Models}
	Electrons and ions both contribute to the stopping of projectiles. However, due to the mass disparity between carbon nuclei and electrons,  $\approx$ 22k $\times$, the interaction between the carbon nuclei and the projectile is small near or above the Bragg peak ($v_p \approx v_e$), and can be neglected. Thus, we only discuss the electronic contribution to the stopping power, ESP. We will simply refer to the stopping power $S_{p}$ in the remainder of the article. For a homogeneous electron liquid / gas (HEG), and within the linear response (Born approximation) regime, the $S_p$ is related to the energy loss function, $ \mathrm{Im}[{-1 \over \varepsilon (G,\omega)} ]$, the projectile charge $Z_p$ and velocity $v_p$:
	
\begin{eqnarray}
S_{p}&=-dE/dx \\ \nonumber
&={ 2 Z_{p}^2 \over \pi v^2_{p} } \int^\infty {dG\over G} \int ^{G\cdot v_{p}} d\omega \: \omega \: \mathrm{Im}[{-1 \over \varepsilon (G,\omega)} ]\:,
\end{eqnarray}
 where $G$ is the electron wavevector and $\omega$ is the angular frequency   \cite{refId0}. 
 
For plasmas of arbitrary degeneracy, the finite temperature Born random phase approximation (BRPA) quantum dielectric can be used \cite{PhysRevA.29.1471},  often referred to as the Maynard-Deutsch (MD) method \cite{refId0}.  The Mermin dielectric approximation can be used to approximate the effects of electron scattering by other electrons or ions.\cite{PhysRevE.76.016405, barriga-carrasco_2008, PhysRevE.94.063203} Electron correlation can also be included through local field corrections, for example using the STLS \cite{PhysRev.176.589, barriga-carrasco_2008} or quantum Monte-Carlo method \cite{doi:10.1063/5.0007476}. 

In addition to the Born dielectric approaches, the Brown-Preston-Singleton \cite{BROWN2005237, doi:10.1063/1.2840134} and Li-Petrasso  \cite{doi:10.1063/1.5114637} models are commonly used. These  and other kinetic theories combine Boltzmann and Lenard-Balescu kinetic equations, with corrections for quantum degeneracy, plasmon excitations, etc \cite{doi:10.1063/1.5095419}. These give analytical expressions, thus efficient enough to be used in-line with hydrodynamics codes \cite{doi:10.1063/1.5119144}, while the MD approach requires non-negligible numerical integration.  Casas and Barriga-Carrrasco incorporated single-atom electronic structure calculations of the mean-excitation energy with Bethe's formula, corrected for the asymptotic limits of the BPRA \cite{refId0, doi:10.1063/1.338141}, to calculate the stopping power of bound electrons \cite{PhysRevE.88.033102, barriga-carrasco_casas_2013}. They then combine this with the BRPA method for the free electrons to extend the calculation from cold gas to plasmas. This method does account for both bound and free electrons, but requires a  $Z^*$ and potentially partial populations of the bound electron orbitals for $T>0$. 

All these models require $Z^*$ in any partially ionized plasmas. We note that the ``local density approximation" has been applied to the calculation of stopping power using an electron density determined from an average atom calculation.\cite{doi:10.1063/1.3420276} This involves averaging the density-dependent BRPA HEG stopping over the calculated range of electron densities.\cite{doi:10.1063/1.3420276} However, this neglects any differences the between responses of an inhomogeneous electron density and a HEG. This approximation is suspect when integrating the response over wavelengths which exceed inter-atomic spacing, or when considering bound electrons. 

\subsection{Deterministic DFT / TD-DFT and Atomistic Electronic Stopping Power}

	In principle, \emph{ab initio} density functional calculations are advantageous in the fact that one does not need to separate and use different models, for different electrons. In practice, for KS TD-DFT of carbon systems,  the K-shell electrons are typically considered frozen in the calculation and treated by a non-local pseudo-potential. Thus, they do not explicitly contribute to the TD response. For the systems here, at $T=10$ eV, the ionization of the K-shell electrons is minimal, and we can include their contribution from a cold-gas model. We choose to use the bound-electron model employed by Casas and Barriga-Carrrasco (CBC method) \cite{barriga-carrasco_casas_2013}. The atomic orbital parameters were determined using the atomic DFT code, with PBE functional \cite{Perdew:1996aa}, written by Hartwigsen for fitting HGH/GTH pseudo-potentials\cite{PhysRevB.54.1703,Hartwigsen:1998aa}. The K-shell contribution is more important at higher velocities and/or higher densities. The use of this model exclusively for the low energy fully-occupied K-shell electrons, with TD-DFT for all other electrons, is a significant improvement compared to combining this CBC model with another model for the free-electrons response and another model for determining $Z^*$. 
	 
\begin{figure}[h]
	\begin{center}
	\includegraphics[width=1.0\columnwidth] {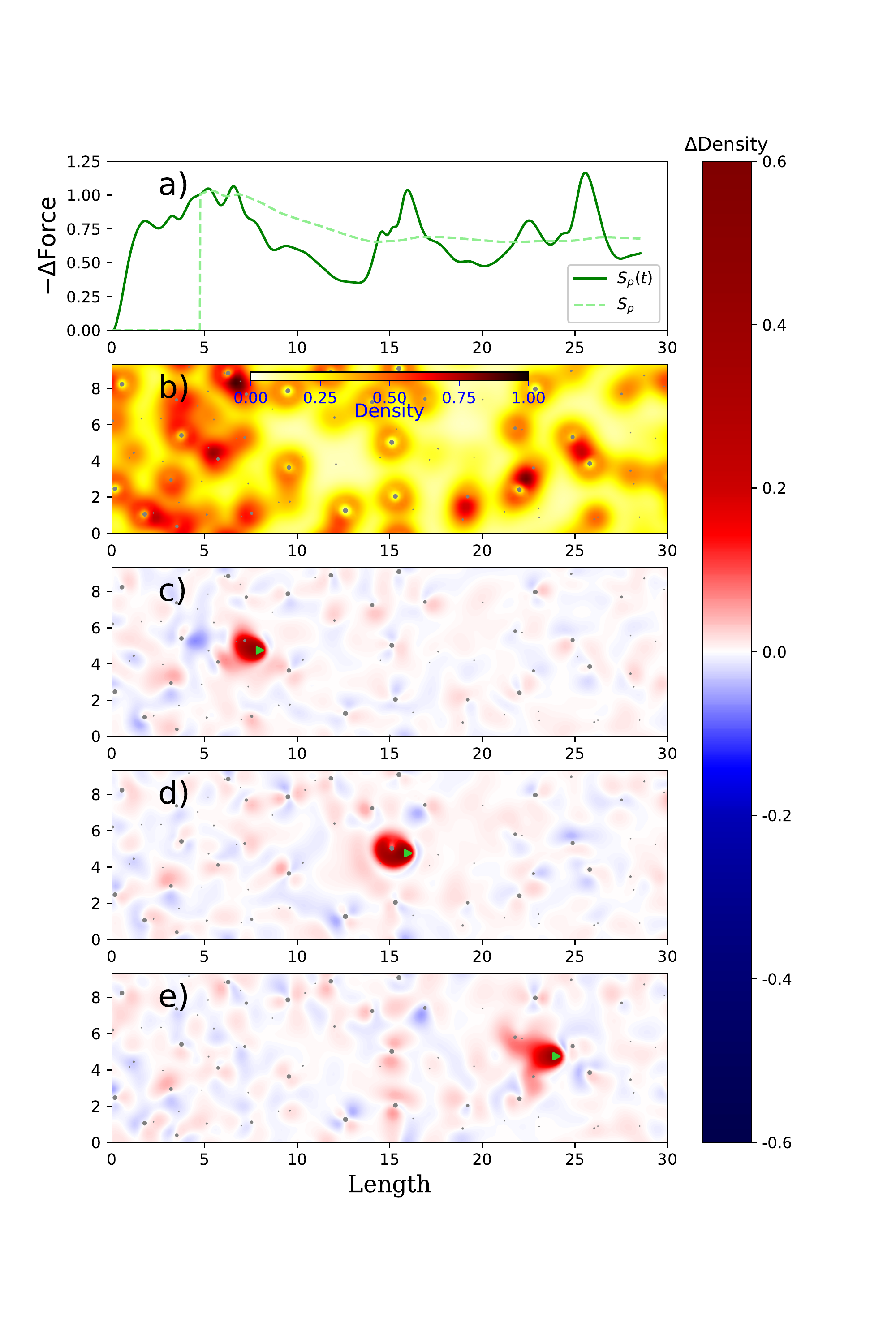}
	\caption{(Color online) Atomistic stopping power simulation, 10 g/cc Carbon at 10 eV, proton projectile with $v_p=2.5$, near Bragg Peak. a) the instantaneous stopping power, $S_{p}(t)$ dark green solid line, and it's running time average, $S_p$ light green dashed line, Eq. \ref{ourSp}. X axis is the time dependent position of projectile. b) a colormap of a 2D slice of the initial electron density, taken at the position of the projectile. Position of atoms is shown as circles with sized defined by distance from the 2D slice. c-e) similar 2D slices, but the difference between the instantaneous and initial density is plotted. The projectile is shown as a green triangle. The colormap for the density difference is shown on the right, and is set to emphasize the small differences near zero. The times are 3.2, 6.4, and 8.6 atomic units. This simulation uses TD-mDFT with $N_{\psi} = 512 $, $N_{\chi} = 64 $ for 256 carbon atoms.}
	\label{snapshot}
	\end{center}
\end{figure}
	
	The ``real time" (RT)-TD-DFT approach has been utilized extensively to calculate the stopping power for condensed phase materials\cite{Caro:2017uw,PhysRevLett.116.043201,Fu_2019,PhysRevLett.108.225504}, single layer graphene\cite{Zhao_2014}, DNA \cite{doi:10.1021/jacs.8b12148}, etc. It has been applied in only a limited number of cases to WDM, due to the computational cost of traditional algorithms.  In RT-TD-DFT, the KS states are propagated via coupled (through the total time-dependent electron density) single-particle TD Schr{\"o}dinger equations. That is,
	 
\begin{eqnarray}
{i\partial\psi_b(t)\over \partial t}&={\hat H_{KS}(t) \psi_{b}(t)}
\\
\hat H_{KS}(t)& =  {-(\vec\nabla+i\vec{A}_{ext})^2\over2} +\int d{\vec R}^{\,3} \frac{\rho({\vec R'})}{\vert {\vec R}- {\vec R'} \vert}
\\\nonumber &\:\:\:+ {\hat V}_{xc}(\rho) +  {\hat V}_{ext} (t)
\\
\label{Mermin}
\textrm{and } \rho(t)&=\sum_b F_D(\varepsilon_b,\mu, T) \vert \psi_{b} (t)  \vert^2 \textrm{,}
\end{eqnarray}
where $\psi_b$'s are Kohn-Sham states, initially eigenvectors of $H_{KS}(0)$ with eigenvalue $\varepsilon_b$,  $\rho$ is the electron density, $F_D$ is the Fermi-Dirac function, $\mu$ is the electron chemical potential, ${\hat V}_{xc}$ is the adiabatic exchange correlation potential, and ${\vec A}_{ext}$ / ${\hat V}_{ext}$ are any other interacting vector / scalar potential, including the potentials due to the bulk nuclei and projectile. Temperature is accounted for through the Mermin formulation \cite{Mermin:1965aa} applied to the initial conditions, Eq. \ref{Mermin}, where $\varepsilon_b$ is the eigenvalue of $H_{KS}(0)$. In order to calculate electronic stopping power our external potential is :

\begin{eqnarray}
\hat{V}_{ext}(t)=&\hat{V}_{ext}(0) \\\nonumber
&+ \theta(t) {\mathcal F^{-1}} [e^{i \{\vec{R}_{0}+{\vec{v}_{p}t \}\cdot \vec{G}}} V_{p}(\vec{G})] \textrm{,}
\end{eqnarray}
where $V_{p}(G)$ is a local all-electron pseudopotential for the projectile nuclei and $\mathcal F^{-1}$ is the inverse Fourier transform. That is, we ``drop" the projectile nuclei into the simulation at the initial time and position, $\vec{R}_0$, propagate it with a constant velocity, $v_p$, while keeping the background ions frozen. This is only a minor approximation, due to the finite simulation time and the velocity difference between the projectile and the bulk nuclei. Practically this allows for the removal of the net-zero contribution to the electronic stopping power from the periodic ground-state density and bulk nuclei. 

We calculate the TD force on the ion and subtract the background force due to the density at t=0,

\begin{eqnarray}
\label{ourSp}
S_{p} (v_p) &=-\langle dE_{KS}/dx (t) \rangle = \int_{t_{min}}^{t_{max}} {dt S_p(v_p, t)\over t_{max}-t_{min}}  \\\nonumber
S_p(v_p, t)&=-\frac{\vec{v}_p}{\vert\vec{v}_p\vert} \cdot \int\: d\vec{G}\:i\vec{G}[\rho^*(\vec{G},t)-\rho^*(\vec{G},0)] \times\\ 
&\:\:\:\:\:\:\:\:\:\:\:\:\:\:\:\:\:e^{i \{\vec{R}_{0}+{\vec{v}_{p}t \}\cdot \vec{G}}} V_{p}(\vec{G})\;.
\end{eqnarray}
Here the force is calculated by the Hellman-Feynman theorem. For periodic boundary conditions and a frozen bulk ion background, the ground-state force: $ -\int\: d\vec{G}\: i{\vec{G}}\rho^*(\vec{G},0) e^{i \{\vec{R}_{0}+{\vec{v}_{p}t \}\cdot \vec{G}}} V_{p}(\vec{G})$, will lead to a net-zero contribution for long time averages, but will produce a highly oscillatory background to the TD force, $ -\int\: d\vec{G}\: i{\vec{G}}\rho^*(\vec{G},t) e^{i \{\vec{R}_{0}+{\vec{v}_{p}t \}\cdot \vec{G}}}$, which makes it difficult to track convergence of the average, unless initial and final times are chosen carefully. We simply subtract this background when calculating $S_p$. 

We discard the beginning portion of the trajectory to eliminate any transient contribution due to the projectile ``drop in". For high velocities, $v_p>v_e$ increasingly larger simulation cells are required to converge the simulation, particularly in the direction of the projectile motion, due to plasmon excitation of increasing wavelength. Thus we use large and elongated (4:1:1 aspect ratios) simulations cells to converge this and any other finite size effects. We systematically test this convergence on representative trajectories.

 In Fig. \ref{snapshot} we show a representative simulation from a single carbon snapshot at 10g/cc and 10 eV with a proton projectile with a velocity of 2.5. Fig \ref{snapshot}-a) shows the calculated instantaneous and running  time-averaged stopping power on the projectile. We have removed the initial $1/8^{\mathrm{th}}$ of the simulation cell from the average. Fig. \ref{snapshot}-b) shows a 2D slice of the initial electron density, taken at the projectile position, with colormap. Dark colors indicate higher electron density. Nuclei positions are shown as dots with size depending on perpendicular distance from the slice. As a passing note, we see transient clusters and diatomics of carbon in the image, based on high electron density seen between nuclei. Fig. \ref{snapshot}-c-e) display the difference in the time dependent and initial density at $t=$ 3.2, 6.4, and 8.6. We see that the stopping power predictably increases when entering the regions of higher electron density near atoms/clusters. The nonlinear colormap is selected to amplify the appearance of small fluctuations.  The projectile position is shown as a green triangle. We see the electron wake, predicted in HEG, is significantly distorted by the background atoms \cite{doi:10.1063/1.3600533,PhysRevA.58.357}.  

\subsection{Stochastic TD-DFT}
 
 \begin{figure}[t]
	\begin{center}
	\includegraphics[width=1.0\columnwidth] {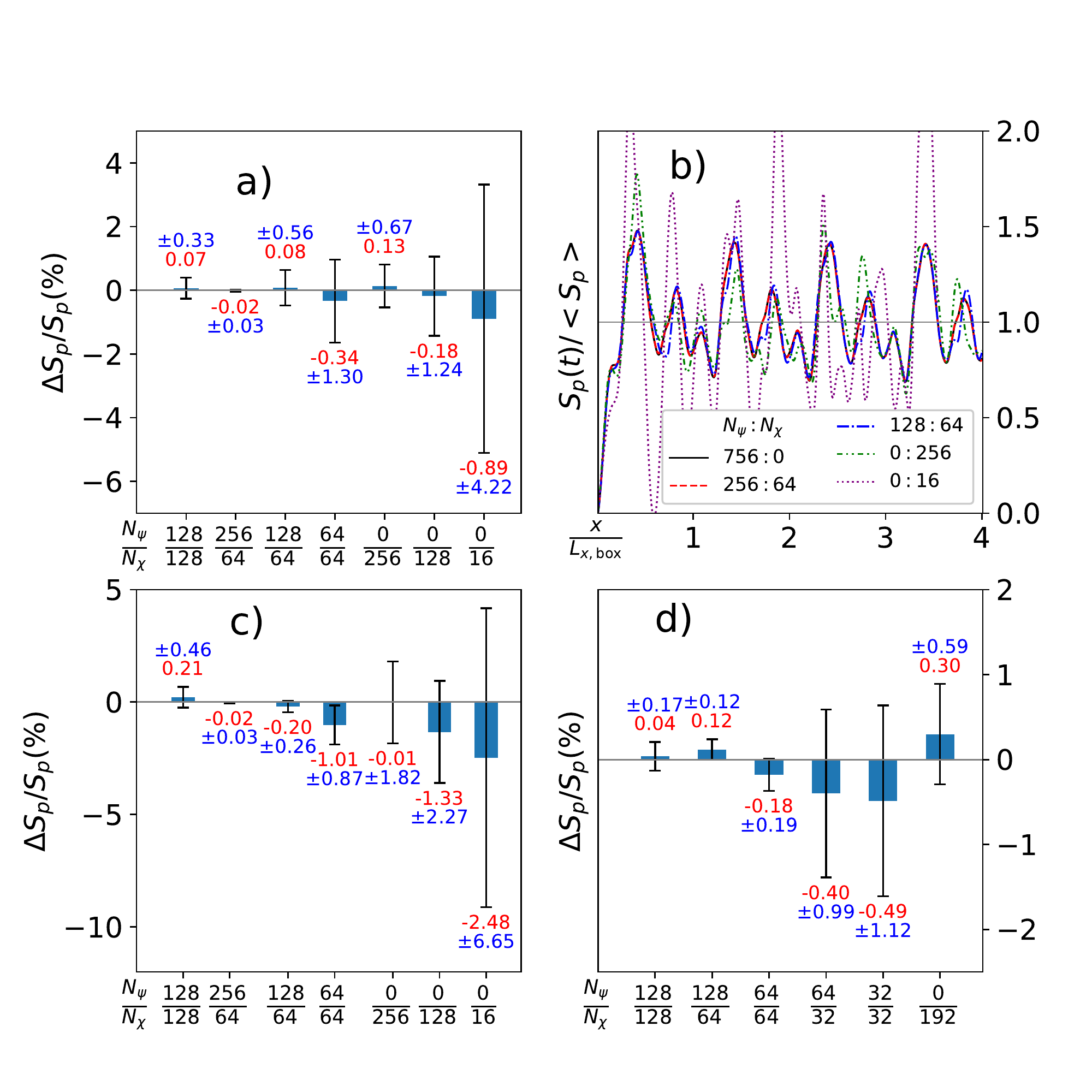}
	\caption{(Color online) Statistics demonstrating convergence of TD-mDFT / TD-sDFT stopping power calculation with respect to the shown number of stochastic vectors, $N_\chi$ and deterministic KS orbitals, $N_\psi$. Calculations with non-zero $N$ for both types are mDFT calculations. a) the difference in proton stopping power, for 10 g/cc 10 eV carbon with $v_p=4.0$, between mDFT/sDFT and dDFT, averaged over 5 repeated calculations for the same ionic configuration, smaller red number and bars. sDFT calculation uses $N_\psi=756$ for 64 atoms. Error bars are $\pm$ one standard deviation over the calculations also shown as $\pm$ blue numbers. b) a representative stopping power trajectory for one calculation but for different  $N_\psi$, and  $N_\chi$. Instantaneous stopping power is scaled by the fully deterministic average. A small cell of 64 atoms is used in order to facilitate converged fully deterministic calculations. The projectile is seen passing through the periodic boundary of the box multiple times. The first two passes are removed from the stopping power time averaging. c) Similar to a), but for $v_p = 2.0$, d) similar to a), but for 3.5 g/cc. For 3.5 g/cc the dDFT references uses $N_\psi=512$ for 32 atoms.}
\label{stats}
\end{center}
\end{figure}

For high temperature systems the number of KS states, $N_\psi$ required to converge the density is $\propto VT^{3/2}$. For the initial self-consistent field calculation this leads to a cost $\propto V^3T^{3}$, due to orthogonalization of the states \cite{doi:10.1063/5.0016538}. For the TD propagation the cost is $\propto V^2T^{3/2}$, due to the nearly linear dependence on $N_\psi$. However, this is still limiting for calculations that need both large system sizes and high temperatures.  

Stochastic density functional theory has emerged over the last decade as a linear scaling method for full KS DFT \cite{Cytter:2018aa,doi:10.1002/wcms.1412}. The method is based on the Hutchison stochastic estimation of the trace of a matrix\cite{doi:10.1080/03610919008812866}, $i.e.$ the density matrix .
Stochastic vectors, $\chi(\vec{R})\equiv {e^{i\phi(\vec{R})}\over\sqrt{d\vec{R}^3}}$, are generated, where $\phi(\vec{R})$ is an  independent random phase, uniformly sampled between $0$ and $2\pi$, for each real-space grid-point. The electron density can be calculated as :
\begin{eqnarray}
\label{Mermin_sdft}
 \rho(t)&= N^{-1}_\chi \sum_\beta^\infty  \chi^*_\beta(t) F_{D}(\hat{H}_{KS}(t=0) ,\mu, T)  \chi_{\beta}(t)  \textrm{.}
\end{eqnarray}
For a finite number of stochastic vectors, there will be stochastic fluctuations in the density. Moreover, the Fermi-Dirac function must now be approximated as a polynomial \cite{Cytter:2018aa}. The cost of this polynomial expansion is $\propto1/T$, while the stochastic precision is increased at a rate of $1/\sqrt{N_\chi}$, as is typical of Monte Carlo methods.  The time dependence of the stochastic vectors is the same as the deterministic KS states, \emph{ i.e.} ${i\partial \chi_\beta(t) \over \partial t}={\hat H}_{KS}(t) \chi_{\beta}(t)$ .  This is a trivial application of the equivalence between Schr{\"o}dinger and Heisenberg pictures. In principle, for systems of sufficient size and temperature, the time-dependent density can be more rapidly converged using this stochastic vectors than the traditional deterministic orbitals. This approach has been applied to the calculation of optical response in clusters \cite{doi:10.1063/1.4905568} , but not, to our knowledge, for WDM or for stopping power. Properties which depend on the integration over the simulation volume and energy domains, \emph{i.e.}, total energies, pressure, etc., can benefit from cancelation of stochastic error over the simulation volume, leading to sublinear scaling of sDFT \cite{Baer:2013aa, doi:10.1063/1.4905568}. However, local quantities such as the electron density or force on a particular atom do not \cite{Chen_2019,Chen_2019_2,Neuhauser_2014}. For stopping power the force on the projectile is calculated as it passes through the simulation volume, thus we expect some cancelation of error, but to a lesser degree than the total energy.  

\subsection{Mixed Stochastic Deterministic TD-DFT}

Recently we developed a mixed-stochastic-deterministic approach for KS-DFT (mDFT) and showed that, for a given precision, the algorithm can lead to significant decrease in the computational time \cite{PhysRevLett.125.055002}. Alternatively, for the same computational time, it can improve the precision and accuracy compared to a purely stochastic calculation. For simulation of ion diffusion coefficients, also in WD carbon, we showed that we can achieve excellent accuracy for significant cost reductions. This is done by ``deflation" of the matrix which is traced stochastically \cite{doi:10.1137/16M1066361, PhysRevLett.125.055002}. That is, we replace the stochastic vectors $\chi$ with the subspace stochastic vectors $\tilde\chi$.
\begin{eqnarray}
\label{mDFT_ortho}
\tilde \chi_{\beta}=  \chi_{\beta}  - \sum_b^{N_b} \psi_b \int d\vec{R}\: \psi^*_b(\vec{R}) \chi_\beta(\vec{R})\textrm{,}
\end{eqnarray}
which are orthogonal to a pre-determined set of KS orbitals (eigenvectors of $\hat{H}_{KS}$).

 The SCF calculation in the Mermin Kohn-Sham method lends itself particularly well to this approach \cite{Mermin:1965aa}. For WDM temperatures the lower energy KS states, where the density of states is lower, but have high occupancy, are calculated deterministically, while the  higher energy orbitals, which have a higher density of states, but have lower individual occupancy, are treated implicitly through the stochastic vectors. The density is given by:
\begin{eqnarray}
\label{mDFT_den}
 \rho(t)&=\sum_b^{N_\psi} F_D(\varepsilon_b,\mu, T) \vert \psi_{b} (t)  \vert^2 + \\\nonumber &N^{-1}_\chi \sum_\beta^{N_\chi}  \tilde\chi_\beta^*(t) F_{D}(\hat{H}_{KS}(t=0) ,\mu, T)  \tilde\chi_{\beta}(t) \\\nonumber
 &=  \sum_i^{N_\psi + N_\chi} W_i  \vert \phi^2_i\vert  \textrm{,}
\end{eqnarray}
 where $\phi$ can be either $\psi$ or $\sqrt{F_{D}(\hat{H}_{KS}(t=0) ,\mu, T)}  \tilde\chi$ with $W_i = F_D(\varepsilon_i,\mu, T)$ or $N^{-1}_\chi$ respectivly. General finite temperature expectation values, including for time-dependent operators, can similarly be expressed as:
 \begin{eqnarray}
\label{mDFT_op}
 \langle \hat{O } \rangle &=  \sum_i W_i   \phi^*_i \hat{O} \phi_i    \textrm{.}
\end{eqnarray}

In Figure \ref{stats}-a we compare the $S_{p}$ calculated from traditional deterministic DFT (dDFT), and various sDFT and mDFT configurations ($N_\psi$ and  $N_\chi$) for a single snapshot of  64 carbon atoms at 10 g/cc carbon at 10 eV, with a proton velocity of 4.0. We take 5 identical trajectories to collect rough statistics on the stochastic sDFT/mDFT error. The percent relative difference between the dDFT $S_{p}$ and the average $S_p$ for each configuration is shown as a bar and red text. The $\pm$  percent relative standard deviation is shown as error bars and blue text.  Fig. \ref{stats}-b shows, for a single representative trajectory, a plot of the $S_{p}(t)$ (Eq. \ref{ourSp}), for some of the configurations, normalized to the TD-dDFT time average $S_{p}$ (Eq. \ref{ourSp}) for the trajectory. The TD-dDFT calculation uses $N_\psi=756$.  Fig. \ref{stats}-c is similar to Fig. \ref{stats}-a, but with a velocity of 2.0. Fig. \ref{stats}-d is similar to Fig. \ref{stats}-a, but for 32 atoms at  a density of 3.5 g/cc, $N_\psi=$ 512 for dDFT,  and different configurations. We see that by increasing either $N_\chi$ or $N_\psi$ we can narrow the deviation to less than 1\%, while using significantly less total orbitals/vectors. For perspective, the standard deviation of the TD-dDFT taken over 5 trajectories with different ionic backgrounds is 6.9\% for the case in Figure a).

\section{Stopping of Carbon}
\begin{figure}[t]
	\begin{center}
	\includegraphics[width=0.9\columnwidth] {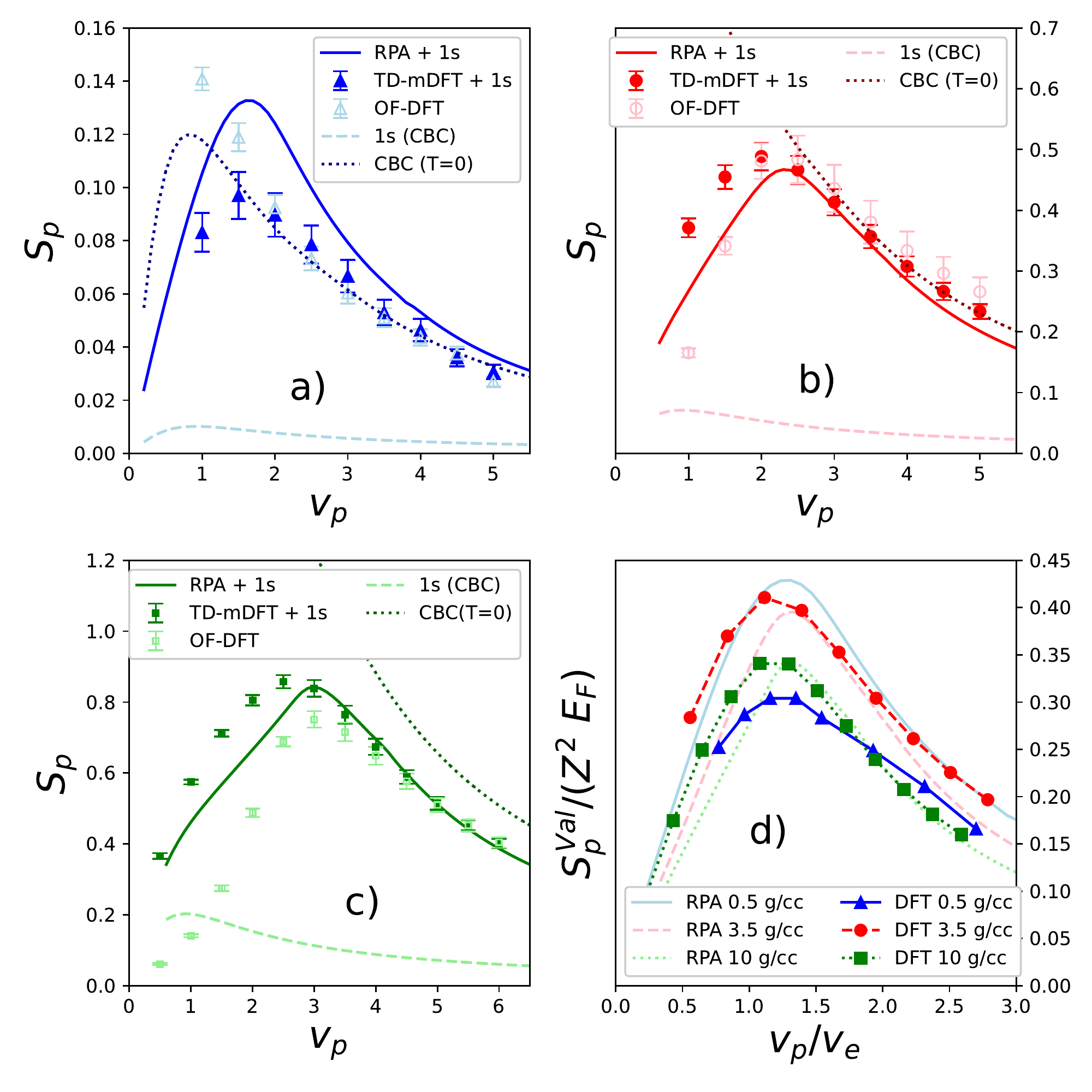}
	\caption{(Color online) a-c) The projectile velocity, $v_p$, dependent stopping power of warm dense carbon at 10 eV and 0.5, 3.5, and 10 g/cc respectively. We compare the TD-mDFT (solid markers)  calculation's, $N_\psi : N_\chi = 16:64, 512:64,$ and $512:64$ respectively, with TD-OF-DFT (open markers),  Born random phase approximation (BRPA, solid line). The BRPA and TD-mDFT are based on 4 electrons per carbon (4EPC), with Casas and Barriga-Carrrasco (CBC) bound electron estimation for the K-shell electrons both added to the results and plotted separately (light dashed lines). The TD-OF-DFT is an all-electron approach. For reference we also plot the CBC result where all 6 electrons are bound, \emph{ i.e.} the result for cold gas. d) we again plot the TD-mDFT results, but without the K-shell contribution and scaled by the relevant plasma parameters,  the Fermi energy, $E_F$,  and the electron velocity $v_e$, again using 4EPC. This gives $E_F=$ 0.29, 1.06 and 2.13 and 0.5, 3.5, and 10 g/cc respectively }
	\label{fig1}
	\end{center}
\end{figure}
\subsection{TD-mDFT results}
	We have simulated the stopping power of carbon in three WDM conditions, 0.5, 3.5, and 10 g/cc, all at $T=10$ eV, using simulation cells of 256, 128 and 256 carbon atoms,  an energy cutoff of 38, 55 and 38, LDA , PBE \cite{Perdew:1996aa} and PBE exchange correlation potentials, respectively. Exchange correlation functionals are implemented through the LibXC package \cite{LEHTOLA20181}. The time step is dependent on energy cutoff for low velocities and projectile velocity at high velocities, ranging from $dt=0.01$ to $0.003$. We utilize the unitary short-iterative Lancoz method to propagate the orbitals/vectors \cite{SCHNEIDER201695}. These parameters are determined by manually checking convergence with respect to sample trajectories, for multiple velocities, at each condition. We employ 4 electrons per carbon (4EPC) HGH non-local pseudo-potentials \cite{PhysRevB.54.1703,Hartwigsen:1998aa}.  These points are indicated as white stars in Fig. \ref{plasma_param} and are representative of a partially ionized blowoff gas, an isochoricaly-heated solid, and shocked compressed dense plasma, respectively. The results are shown in Figure \ref{fig1}. The low density, high proton velocity cases (0.5 g/cc $v_p$ = 4-5) was recently compared to experimental measurements and demonstrated superior agreement to standard models.\cite{PPR:PPR393148} Here, we compare to the BRPA result with 4EPC, and the TD-OF-DFT result. We have added the 1s core electron contribution, estimated by the CBC method (also plotted), to the BRPA and TD-mDFT results. The TD-OF-DFT utilizes a 6 electron local pseudo-potential, thus no core contribution is added.  Additionally, we plot the CBC result assuming all electrons are bound (i.e. for $T\approx 0$). 

The BRPA agreement with the TD-mDFT is better for the two denser systems, while the CBC result agrees reasonably well for the low density case, but overestimates in the dense systems. The TD-OF-DFT provides statistically quantitative agreement in the velocities above the Bragg peak, but either overestimates, or underestimates the result below the Bragg peak, depending on the conditions.  This is consistent with our previous OF/KS comparison.  To compare the valance contribution across the densities, we plot the scaled TD-mDFT stopping power (with no K-shell addition) for scaled velocities. This valence stopping power, $S^{val}_{p}$ is scaled by the Fermi energy, $E_F$ and the square of the projectile charge, $Z$, while the velocity is scaled by $v_e$ (Eq. \ref{ve} ). We used 4EPC for calculating $v_e$ and $E_F$, considering the tightly-bound 1s electrons as separated from the plasma. We see that the position of the Bragg peak is consistent with this basic plasma parameter, across densities, however the magnitude of the 0.5 g/cc stopping breaks the monotonic trend one would expect simply based on BRPA. This would indicate a breakdown of the 4EPC plasma assumption.

\subsection{Electrical Conductivity, relaxation times, and Mermin Dielectric}
\begin{figure}[t]
	\begin{center}
	\includegraphics[width=0.75\columnwidth] {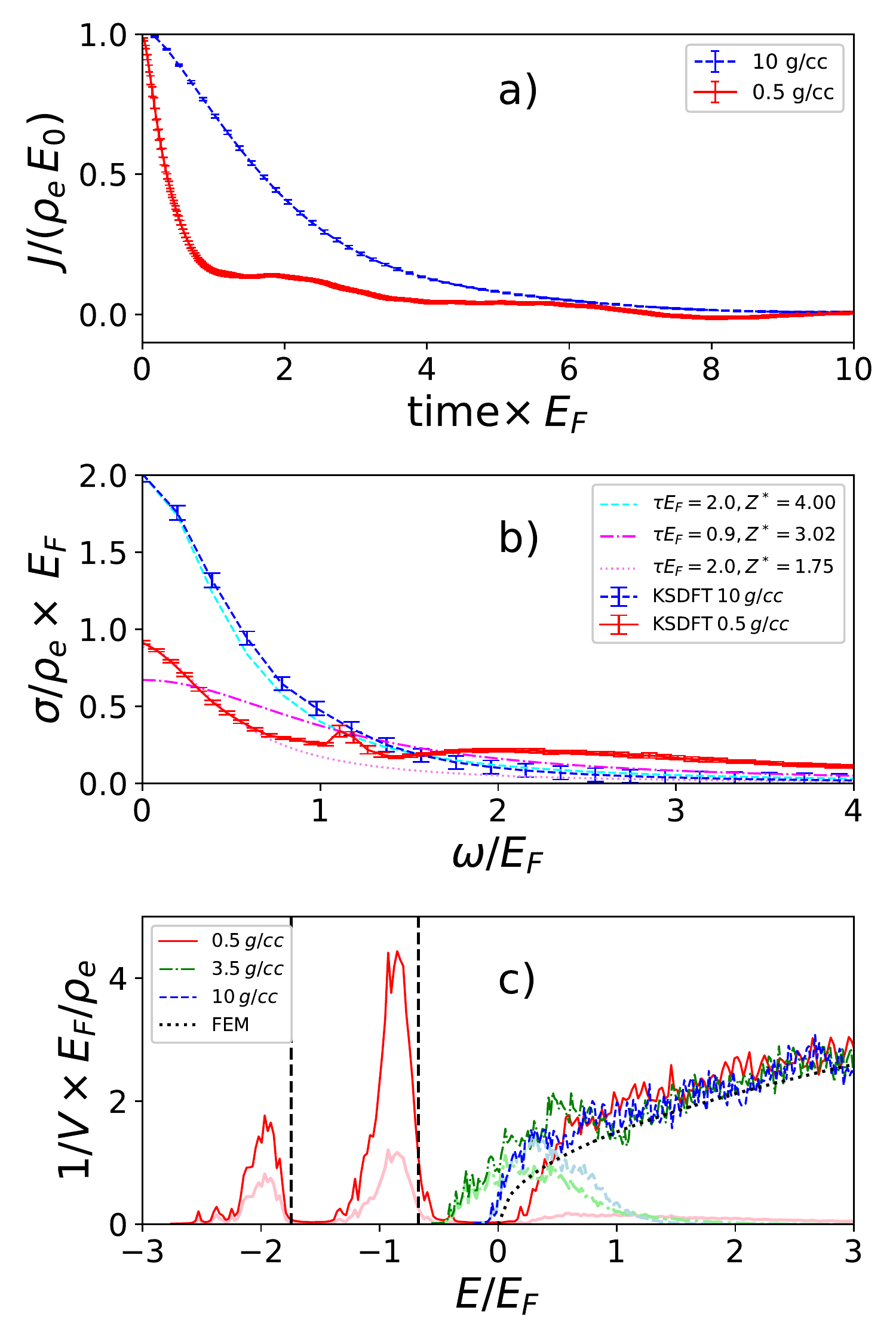}
	\caption{(Color online) a) The current density, scaled by the electric field pulse strength and electron density vs the time (multiplied by the $E_F$). b) The conductivities, \emph{i.e.} Fourier transforms of the top panel, multiplied by $E_F$. Multiple Drude Fits are also plotted (Eq. \ref{Drude}). c) The density of states (DOS, thin dark lines) and occupied DOS(thick light lines) for the three densities. Both densities of states are scaled by the number of electrons in the system and multiplied by the $E_F$, the energy axis is scaled by $E_F$. The DOS of a non-interacting free electron gas (depends only on $\rho_e$ and $E_F$) is plotted for reference. $E_F=$ 0.29, 1.06 and 2.13 for 0.5, 3.5, and 10 g/cc respectively.}
	\label{fig3}
	\end{center}
\end{figure}

	To understand the physical differences between the low and high density cases, which extend beyond the degeneracy and coupling effects of the plasma, we can look to the conductivity and density of states. Previous attempts to calculate stopping power based on fitting dielectric functions to experimental or theoretical, in the optical $G \to 0$ limit, has been reported previously \cite{Ashley_1991, Heredia_Avalos_2007}. In this limit the dielectric and conductivity are straightforwardly related to one another, but the conductivity is easily calculated from RT-TD-DFT, including TD-mDFT, by subjecting the system to an instantaneous in time, uniform in space, electric field pulse. That is:
\begin{eqnarray}
\label{current}
A_{x, ext}(t)&= \Theta(t-t_0) E_0 \\
\sigma(\omega)&= \int dt e^{-i\omega t} J_{x}(t)/E_0 \\
J_{x}(t)&= Re\big\{\sum_i \phi^*(t) [\hat{H}_{KS}(t),R_x] \phi(t) \big\} \:,
\end{eqnarray}	
where $E_0$ is the electric field strength, $\Theta$ is the Heaviside function, $\sigma$ is the conductivity, $\vec{J}$ is the current density, and  $[.,.]$ is the commutator. In Figure \ref{fig3}-a we plot the current, scaled by the field strength and electron density, $\rho_e$. We scale the time by $E_F$. $\rho_e$ and $E_F$ are again calculated assuming 4EPC.   We calculated the current by TD-mDFT using $N_\psi=256$, $N_\chi=64$, $E_0=0.5$, and the same atomic cells as was used for the stopping power calculations. We tested the field strength as low as $E_0=0.005$ to insure the field is within the linear response regime. In Fig. \ref{fig3}-b we plot the corresponding conductivity scaled by the density and multiplied by the Fermi energy vs the angular frequency, $\omega$, scaled by the Fermi energy. Error bars are due to sampling over both stochastic vectors and ion snapshots. 

One of many possible choices to determine a $Z^*$ for carbon under these conditions, is to fit the low frequency real conductivity to a Drude model:
\begin{eqnarray}
\label{Drude}
\sigma (\omega) = {n_a Z^* \tau \over 1 + (\omega \tau)^2}
\end{eqnarray}	
where $n_a$ is the atomic number density. Despite scaling the current and conductivities, the two systems show the dramatic difference between the hot ionized gas and compressed plasma systems. In the latter, the conductivity is well-fit, globally, by a Drude model, indicating that all 2s and 2p electrons behave similar to nearly-free electrons, $i.e.$ $Z^*\approx 4$ and $\tau \approx 2/E_F$, see light dashed line.  In the prior, higher frequency oscillations appear at $\omega \approx 0.3$, $0.35$, and $0.62$. These roughly correspond to $2p\to$ continuum, $2s\to2p$ (facilitated by partial ionization), and $2s\to$ continuum, though broadened and shifted by interaction between ions. The $2s\to2p$ peak is significantly sharper than the excitations to continuum. Due to these excitations the conductivity is not well-fit, globally, by a Drude model, see dash-dot line. However, we can fit the low frequencies, $\omega < 0.25 $, to a Drude with $Z^*\approx 1.75$ and $\tau \approx 2/E_F$, see dotted line.

 Additionally, we calculate the density of states (DOS) and Occupied DOS (ODOS), for one snapshot.

\begin{eqnarray}
\label{DOS}
DOS(E)&\approx 2 \int dt e^{i 2 \pi [E t-i {\gamma \over 2}] t} Tr \{e^{-i \hat{H}_{KS} t}\}  \\ \nonumber
&=2\:\:Tr \{\tilde\delta(E-\hat{H}_{KS})\}\\
ODOS(E)&=Tr \{ F_{D}(H_{KS}(t=0)) \tilde\delta(E-\hat{H}_{KS})\}
\end{eqnarray}	
\begin{figure}[t]
	\begin{center}
	\includegraphics[width=0.9\columnwidth] {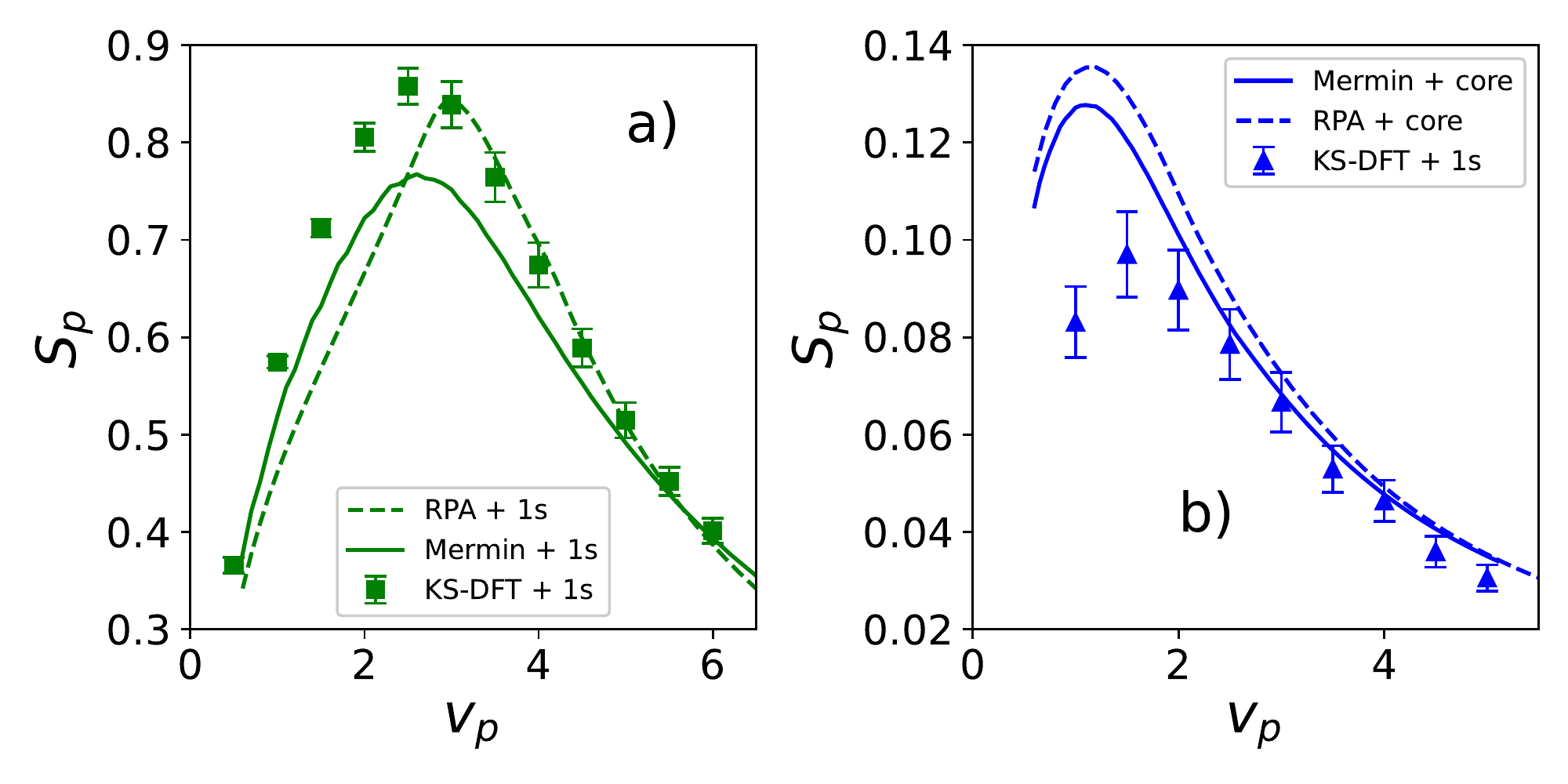}
	\caption{(Color online) Projectile velocity dependent stopping power. The TD-mDFT results are repeated from Fig. \ref{fig1}, Mermin dielectric plus CBC, solid line, assuming $Z^*=4.0$ and $1.75$ for 10g/cc (a) and 0.5 g/cc (b) respectively and $\tau=2.0/E_F$. BRPA plus CBC, dashed line, assuming  $Z^*=4.0$, $Z_{2s}=0$, and $Z_{2p}=0$ for 10 g/cc and $Z^*=1.45$, $Z_{2s}=0.94$, and $Z_{2p}=1.62$ for 0.5 g/cc. }
	\label{fig4}
	\end{center}
\end{figure}
where $\tilde\delta$ is a sharp Lorentzian approximation to a delta function with width $\gamma$. The DOS and ODOS for all three densities is plotted in Fig. \ref{fig3}-c. For reference the continuum of states for a non-interacting free electron gas, $DOS(E)=\frac{3}{2}\frac{\rho_e}{E_F}(\frac{E}{E_F})^{1/2}$, is shown as a dotted line.  For the 0.5 g/cc case the 2s and 2p peaks are distinct from the continuum band, with only small shift downward from their single atom, $T=0$, energies, shown as dashed vertical lines. While for the 3.5 and 10 g/cc cases the continuum has dropped and mixed with the atomic states.  Qualitatively this makes it clear why the BRPA disagrees with the KS-DFT significantly for the 0.5 g/cc, but works reasonably well for the solid density and dense plasma cases.

	We can attempt to improve the stopping power models by using the Drude model parameters, $Z^*$ and $\tau$,  extracted from our conductivity calculations.  We can include the scattering time, $\tau$, into the Mermin formulation \cite{PhysRevE.76.016405} for the dielectric function: 
\begin{eqnarray}
\label{MerminSP}
\varepsilon_M(G,\omega) = 1+ \frac{(\omega +  \frac{i}{\tau})[\varepsilon(G, \omega +  \frac{i}{\tau})-1]}{\omega + \frac{i}{\tau} [\varepsilon(G, \omega +  \frac{i}{\tau})-1]/[\varepsilon(G,0)]-1]},
\end{eqnarray}
 where $\varepsilon$ is the RPA response function, and use the $Z^*$ to adjust the Fermi Energy. 
 
 For the 10g/cc case, $Z^*$ from the conductivity fit is still 4EPC, so the only effect is the inclusion of scattering time. In  Fig. \ref{fig4}-a, we see that the inclusion of the finite scattering time increases the stopping for low velocities, in better agreement with TD-mDFT, but lowers it for higher velocities, where the agreement with BRPA was already quantitative.  An overall improvement is not clear. For the 0.5 g/cc case, $Z^*$ should be reduced due to the 2s and 2p electrons which remaining bound. Reducing $Z^*$ then requires us to include a larger contribution from the CBC method for the bound electrons. Fitting the conductivity does not provide a value for the population of each bound state, only the total $Z^*$. We can approximate $Z_{2s}$ and $Z_{2p}$ by integrating over the two peaks in the $ODOS(E)$, \emph{i.e.} from $-3\:E_F$ to $-1.6\: E_F$ and from $-1.6\:E_F$ to $0$, respectively. This yields a $Z_{2s}=0.94$, $Z_{2p}=1.61$, and $Z^*=1.45$. We assume that the deeply bound 1s orbitals (K-shell) are still fully populated.  
 
 This lower $Z^*$, compared to the Drude fit, is due to partial ionization of the 2s and 2p bands, which can have intra-band transitions that contribute to low frequency $\sigma$, and hence a higher $Z^*$ predicted for the Drude fit. These transitions are not allowed for single atoms. However, they are allowed for disordered condensed phases. Inconsistency between the $ODOS$ and $\sigma$ derived $Z^*$, even from the same electronic structure, but with different definitions, points to the limitation of the concept.  Thus to compare to KS results, we calculate two CGC plus dielectric curves, one with the $Z$'s from the $ODOS$ and using BRPA dielectric (infinite $\tau$), and one with Drude $Z^*$ and the Mermin RPA. In the latter case the ionization is assumed to come fully from the highest energy occupied orbital (the 2p) in the CGC calculation. We see that in both cases the agreement is improved for high velocities, but neither predicts the decreased magnitude near and below the Bragg peak seen in TD-mDFT. 
 \section{Conclusion}
 	We have presented a new approach for \emph{ab initio} time-dependent Kohn-Sham density functional theory, based on mixed deterministic and stochastic methods. This method obtains the efficiency required to perform stopping power calculations at many eV temperatures, with the large box sizes required to overcome finite size effects. We can utilize these high fidelity atomistic results to interrogate more easily calculated stopping power models. Here we have focused on using the BRPA and Mermin dielectric functions for the nearly free electron contribution with the bound electron contribution from the CGC method. Unfortunately,  quantitative agreement is not significantly improved by including parameters taken directly from the \emph{ab initio} method. This could indicate that improvements to the models are needed, such as improved dielectric functions \cite{barriga-carrasco_2010, PhysRevE.97.023202} or bound state energies derived from finite temperature average atom simulations with density effects included \cite{GILL201733}. It could indicate a fundamental limitation of the separation between bound and free electrons in the calculation of stopping power of condensed phases, or limitations of the BRPA and Mermin dielectric functions for dissorederd plasmas. Additional studies will certainly be required. 
	
	For now, this motivates continued use of TD-DFT to validate or invalidate models in different parameter regimes. Tuning of tight-binding models is another potential route to achieve efficiency and accuracy \cite{Race_2013, Mason_2012} .  Alternatively, one could seek to improve the TD-OF-DFT results through novel kinetic energy functionals. We note that, other than TD-OF-DFT, all calculations presented here have used the same method to calculate the K-shell contribution to the stopping power. Future work will involve calculation of the 1s stopping using TD-KS-DFT and an all-electron pseudo-potentials.   

\section{Acknowledgment}

This work was supported by the U.S. Department of Energy through the Los Alamos National Laboratory (LANL). Research presented in this article was supported by the Laboratory Directed Research and Development program of LANL, under project number 20210233ER, and Science Campaign 4. This research used computing resources provided by the LANL Institutional Computing and Advanced Scientific Computing programs. Los Alamos National Laboratory is operated by Triad National Security, LLC, for the National Nuclear Security Administration of U.S. Department of Energy (Contract No. 89233218CNA000001). This work was supported by the Department of Energy National Nuclear Security Administration Award Number DE-NA0003856 through Laboratory for Laser Energetics, at University of Rochester.

\section{References}
\bibliographystyle{iopart-num}
\bibliography{mDFT}
\end{document}